\documentclass[oldversion]{aa}  
\usepackage{graphicx,float}
\usepackage{latexsym,amsmath,amssymb}
\usepackage{natbib}
\usepackage{changebar}
\usepackage{subfigure}
\usepackage{txfonts}
\usepackage{rotating}
\usepackage{longtable, lscape}
\usepackage{color, colortbl}
\usepackage{multirow}

\definecolor{Gray}{gray}{1}

\usepackage{natbib}
\bibpunct{(}{)}{;}{a}{}{,}

\def\apjl{ApJ}%
\def\apjs{ApJS}%
\def\ao{Appl.~Opt.}%
\def\aap{A\&A}%
\begin{document}

   \title{Multiwavelength campaign on Mrk 509}

   \subtitle{XIII. Testing ionized-reflection models on Mrk 509}

   \author{R. Boissay
          \inst{1},
          S. Paltani
          \inst{1},
G. Ponti	
\inst{2},
S. Bianchi
\inst{3},
M. Cappi
\inst{4},
J.S. Kaastra
\inst{5,6},
P.-O. Petrucci
\inst{7},
N. Arav
\inst{8},
G. Branduardi-Raymont
\inst{9},
E. Costantini
\inst{5},
J. Ebrero
\inst{5},
G.A. Kriss
\inst{10,11},
M. Mehdipour
\inst{5,9},
C. Pinto
\inst{5},
K.C. Steenbrugge
\inst{12,13}
          }

   \institute{ Department of Astronomy, University of Geneva, ch. d'\'Ecogia 16, 1290 Versoix, Switzerland  \and Max-Planck-Institut f\"{u}r extraterrestrische Physik, Giessenbachstrasse, D-85748 Garching, Germany \and Dipartimento di Matematica e Fisica, Universit\`{a} degli Studi Roma Tre, via della Vasca Navale 84, 00146 Roma, Italy \and INAF-IASF Bologna, Via Gobetti 101, 40129 Bologna, Italy \and SRON Netherlands Institute for Space Research, Sorbonnelaan 2, 3584 CA Utrecht, the Netherlands \and Sterrenkundig Instituut, Universiteit Utrecht, P.O. Box 80000, 3508 TA Utrecht, the Netherlands \and UJF-Grenoble I / CNRS-INSU, Institut de Plan\'{e}tologie et d'Astrophysique de Grenoble (IPAG) UMR 5274, Grenoble, F-38041, France \and Department of Physics, Virginia Tech, Blacksburg, VA 24061, USA \and Mullard Space Science Laboratory, University College London, Holmbury St. Mary, Dorking, Surrey, RH5 6NT, UK \and Space Telescope Science Institute, 3700 San Martin Drive, Baltimore, MD 21218, USA \and Department of Physics and Astronomy, The Johns Hopkins University, Baltimore, MD 21218, USA \and Instituto de Astronom\'{i}a, Universidad Cat\'{o}lica del Norte, Avenida Angamos 0610, Casilla 1280, Antofagasta, Chile \and Department of Physics, University of Oxford, Keble Road, Oxford OX1 3RH, UK 
             }
   \authorrunning{ R. Boissay et al.}
   \titlerunning{Testing ionized-reflection models on Mrk 509}
    \date{Received; accepted}

\abstract{Active Galactic Nuclei (AGN) are the most luminous persistent objects in the universe. The X-ray domain is particularly important as the X-ray flux represents a significant fraction of the bolometric emission from such objects and probes the innermost regions of accretion disks, where most of this power is generated.
An excess of X-ray emission below $\sim$ 2 keV, called soft-excess, is very common in Type 1 AGN spectra. The origin of this feature remains debated. Originally modeled with a blackbody, there are now several possibilities to model the soft-excess, including warm Comptonization and blurred ionized reflection.
In this paper, we test ionized-reflection models on Mrk 509, a bright Seyfert 1 galaxy for which we have a unique data set, in order to determine whether it can be responsible for the strong soft-excess.
We use ten simultaneous XMM-Newton and INTEGRAL observations performed every four days. We present here the results of the spectral analysis, the evolution of the parameters and the variability properties of the X-ray emission.
The application of blurred ionized-reflection models leads to a very strong reflection and an extreme geometry, but fails to reproduce the broad-band spectrum of Mrk 509. 
Two different scenarios for blurred ionized reflection are discussed: stable geometry and lamp-post configuration. In both cases we find that the model parameters do not follow the expected relations, indicating that the model is fine-tuned to fit the data without physical justification. A large, slow variation of the soft-excess without counterpart in the hard X-rays could be explained by a change in ionization of the reflector. However, such a change does not naturally follow from the assumed geometrical configuration.
Warm Comptonization remains the most probable origin of the soft-excess in this object. Nevertheless, it is possible that both ionized reflection and warm Comptonization mechanisms can explain the soft-excess in all objects, one dominating the other one, depending on the physical conditions of the disk and the corona.}


   \maketitle
   %______________________________________________________________
   
\section{Introduction}

In non-obscured Active Galactic Nuclei (AGN), most of the radiation is emitted in the optical-UV and the X-ray energy bands.
In the optical-UV band, the emission is characterized by the ``big blue bump", present from about 10 nm to 0.3 $\mu$m \citep{Sanders1989,Bregman1990,Zhou1997}. This emission is thought to come from an optically thick accretion disk \citep{ShakuraSunyaev1973}.

The X-ray spectrum of Seyfert galaxies is typically characterized by a power-law continuum with reflection features, absorption and often an excess in the soft X-rays \citep{Halpern1984,Turner1989}.
The power-law emission is believed to be due to Comptonization of the UV photons coming from the disk by the energetic electrons in a corona surrounding the disk \citep{Blandford1990,Zdziarski1995,Zdziarski1996,Krolik1999}. 
As a high-energy cut-off is detected in a range from about 80 to 300 keV in about 50\% of Seyfert galaxies, a thermal distribution is preferred to a non-thermal one \citep{Gondek1996,Matt2001,Perola2002}. 

The strongest signatures of the reflection component are a ``reflection hump" around 30 keV and an iron FeK$\alpha$ fluorescence line in between 6 and 7 keV depending on the iron ionization state. Reflection is associated with the reprocessing of the primary continuum by material either close to the central black hole, in the accretion disk \citep{George1991,Matt1991} or more distant, for example in a torus
\citep{Antonucci1993,Jaffe2004,Meisenheimer2007,Raban2009}. It can also be produced either in the narrow or broad line regions \citep{Ponti2013}.

Absorption from material either in the vicinity of AGN or in the host galaxy is generally observed in the X-ray spectra of Seyfert 1 and 2 galaxies. If the absorbing material is often photoionized, it is referred to as ``warm absorber" \citep{George1998}.

More than 50\% of Seyfert 1 galaxies show the presence of a soft X-ray excess called soft-excess \citep{Halpern1984,Turner1989}, a soft X-ray emission below $\sim$ 2 keV in excess of the extrapolation of the hard X-ray continuum. \cite{Piconcelli2005} and \cite{Bianchi2009} even found that the fraction of AGN with soft-excess reaches about 100\%. The discovery of this component was made thanks to the HEAO-I  \citep{Singh1985} and EXOSAT \citep{Arnaud1985} missions in the 80's, but its nature is still uncertain. 
It was first thought to arise from the hottest part of the accretion disk \citep[e.g.][]{Arnaud1985,Pounds1986}, but this hypothesis was invalidated by the facts that the temperature of the soft-excess (0.1-0.2 keV) is much too high to be explained by the standard accretion disk model around a supermassive black hole, and that it does not vary, as expected, with the mass of the black hole \citep{GierlinskiDone2004}. 

Another possible explanation is ``warm" Comptonization: up-scattering of seed disk photons in a Comptonizing medium which has a temperature of about 1 keV (e.g. in NGC 5548 -- \citealt{Magdziarz1998}; RE J1034+396 -- \citealt{Middleton2009}; RX J0136.9--3510 -- \citealt{Jin2009}; Ark 120 -- \citealt{Matt2014}; and 1H 0419--577 -- \citealt{DiGesu2014}). 
This Comptonization model is supported by the fact that strong similarities have been found in the optical-UV and soft X-ray variability, suggesting a correlation between these emissions, in agreement with inverse Compton processes \citep{Edelson1996,Mehdipour2011}. 
\cite{Walter1993} used a sample of 58 Seyfert 1s observed by ROSAT and IUE and found a spectral shape correlation (i.e. an amplitude correlation) indicating that the soft-excess could be the high-energy tail of the "big blue bump" component observed in UV, as objects with a large UV bump do also show a strong soft-excess. 
\cite{Edelson1996} found a  correlation between variabilities in X-rays, UV and optical bands in NGC 4151. \cite{Marshall1997} studied NGC 5548 in the extreme UV band using EUVE and noticed that the EUV and UV/optical variations of the light curves are simultaneous. They also found that the shape of the EUVE spectrum is consistent with that in UV and soft X-rays.

However, the warm Comptonization model still does not explain why the shape of the soft-excess does not appear to vary with the black-hole mass \citep{GierlinskiDone2004}.
An alternative explanation is that the soft-excess is linked to atomic processes. The soft-excess could be the signature of strong, relativistically smeared, partially ionized absorption in a wind from the inner disk. 
\cite{GierlinskiDone2004} applied this model on PG 1211+143.
If a totally covering absorber is assumed, the problem of this model is that it needs quite extreme values of the model parameters to account for the observed smooth soft-excesses, in particular a very large smearing velocity, which is not attainable in models of radiatively driven accretion disk winds with typical physical parameters \citep{Schurch2007,Schurch2009}. This problem does not apply anymore in the case of a partial-covering scenario.
In Mrk 766, for example, the principal components analysis and spectral variability of a long XMM-Newton observation can be explained by ionized absorption partially covering the continuum source \citep{Miller2007}. 
\cite{Turner2009} present a review on X-ray absorption and reflection in AGN, showing that partial-covering absorption can explain spectral curvature and variability at low energy.

Another interpretation is that the soft-excess is the result of ionized reflection in a relativistic disk, which blurs all emission lines.
An ionized-reflection model, calculated for an optically-thick atmosphere of constant density illuminated by radiation of a power-law (called \textit{reflionx} in XSPEC; \citealt{RossFabian2005}), has been successfully applied by \cite{Crummy2006} on 22 Type-1 PG quasars and 12 Seyfert 1 galaxies and by \cite{Zoghbi2008} on Mrk 478 and EXO 1346.2+2645. 
This model has also been applied in MCG-6-30-15 by \cite{Vaughan2004} and NGC 4051 by \cite{Ponti2006}, explaining the spectral shape as well as the variability.
The spectral and timing analysis of 1H 0707-495 \citep{Fabian2009} provides strong evidence of emission from matter close to a rapidly spinning black hole. This object shows the presence of a broad iron K line whose width and shape is a signature of strong gravity and spin of the black hole. Thanks to the high iron abundance, the iron L line is also detectable and a lag of about 30 seconds with the direct X-ray continuum could be measured. This lag is an evidence of
reverberation processes. X-ray reverberation time delays have also been detected in MCG--5--23--16 and NGC 7314 by \cite{Zoghbi2013} using the iron K$\alpha$ emission lines.  \cite{Kara2013} also found iron K lags in Ark 564 and Mrk 335. Soft X-ray reverberation lags have been found in ESO 113-G010 by \cite{Cackett2013} and PG 1211+143 by \cite{DeMarco2011}. Overall, soft lags have been found in more than 15 sources \citep{DeMarco2013}.
Blurred ionized reflection has also been tested, as well as double Comptonization and ionized absorption by a high velocity material, in Mrk 509 and Mrk 841, using average broad-band Suzaku data \citep{Cerruti2011}. This model seems to correctly describe Mrk 509 soft-excess, but underestimates the broad iron emission line.

In this paper, we focus on testing blurred ionized-reflection models in an object for which we have a unique data set: Mrk 509. 
We want to check whether ionized reflection could be a viable alternative explanation to the warm Comptonization to explain the soft-excess. In order to test these ionized-reflection models, we use the Mrk 509 XMM-Newton and INTEGRAL campaign and its 10 simultaneous observations \citep{Kaastra2011}.
We first study the variability at different time scales and then we test blurred ionized-reflection models on the ten observations in order to investigate the evolution of the \textit{reflionx} parameters.

The paper is organized as follows. In Sect. \ref{2}, we first introduce the properties of Mrk 509 and the XMM-Newton/INTEGRAL campaign of observation that we use. In Sect. \ref{3}, we show the results of the study on variability at different time scales. Section \ref{4} presents the ionized-reflection models that we use to fit our data, as well as the results on the average spectrum and for each observation. In Sect. \ref{5}, we show the evolution of \textit{reflionx} parameters, taking into account two cases: the case of a stable geometry with a constant reflection factor and the case of a varying reflection factor. All these results are discussed in Sect. \ref{6} and we state the conclusions in Sect. \ref{7}.

%%%%%%%%%%%%%%%%%%%%%%%%%%%%%%%%%%%%%%%%%%%%%%%%%%%%%%%%%%%%%%%%%%%%%%

\section{Mrk 509 -- XMM-Newton/INTEGRAL campaign}
\label{2}

\begin{figure} [!h]
\resizebox{\hsize}{!}{\includegraphics[trim = 30mm 30mm 18mm 35mm, clip, angle=-90]{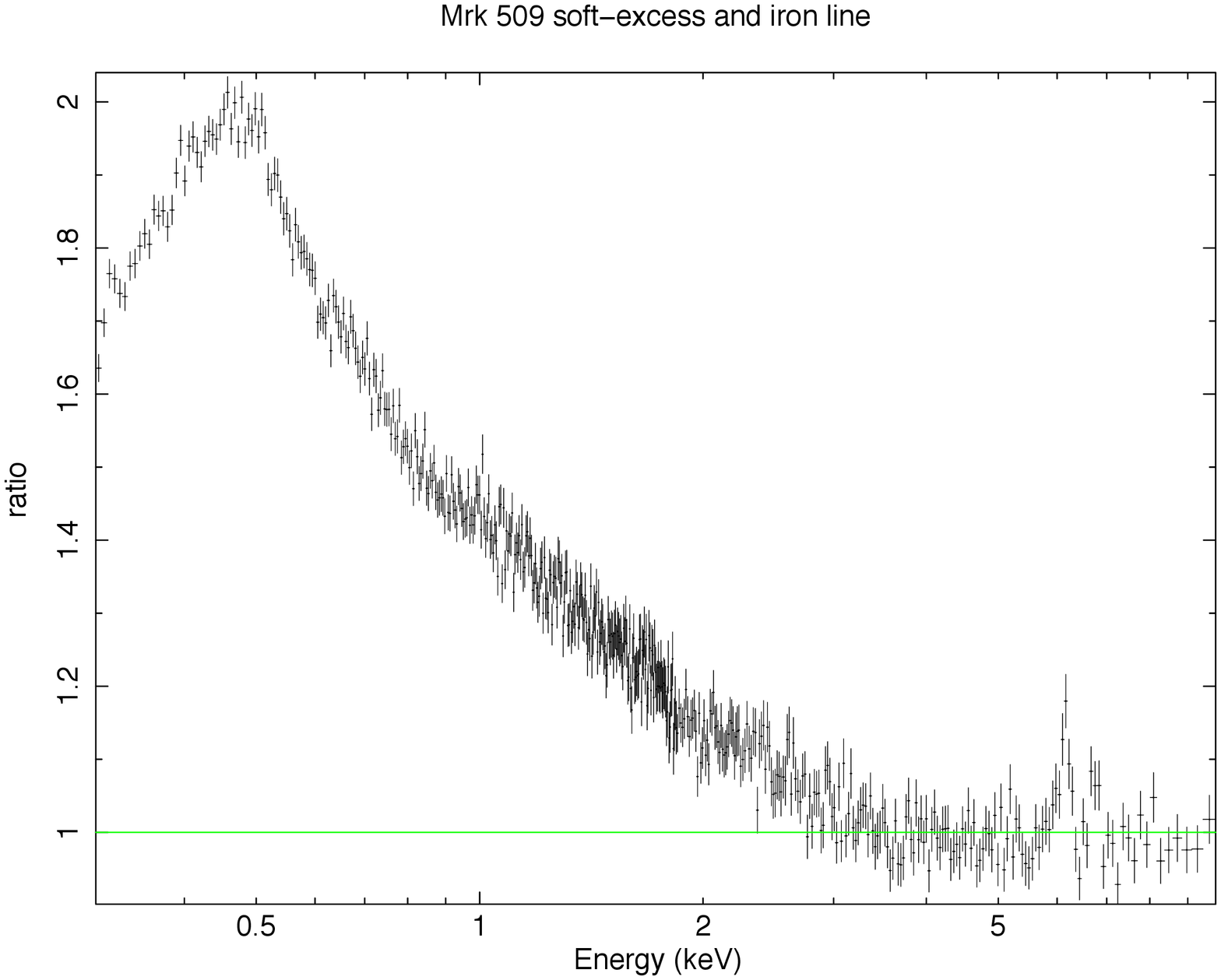}} 
\caption{Ratio to the power-law fit between 3 and 10 keV showing the presence of a strong soft-excess below 2 keV and a resolved iron line at 6.4 keV in the average spectrum of Mrk 509 (431 ks of XMM-Newton exposure).}
\label{RatioSE}
\end{figure}

Mrk 509 is a bright Seyfert 1 galaxy. It has a redshift of 0.034 \citep{Huchra1993}. Its black-hole mass $\text{M}_{\text{BH}}=1.43 \times 10^{8} \text{M}_{\odot}$ has been determined  using the reverberation mapping method \citep{Peterson2004}. This object presents a strong soft-excess, as shown in Fig. \ref{RatioSE}.

Mrk 509 has been one of the first AGN to be studied, thanks to its brightness, with a flux between 2 and 10 keV of $F_{\text{2-10 keV}}=2-5 \times 10^{-11} \text{erg cm}^{-2}\text{s}^{-1}$ \citep{Kaastra2011}. Its bolometric luminosity is  $\text{L}_{\text{Bol}}=1.07 \times 10^{45} \text{erg s}^{-1}$ \citep{Woo2002} and it has an X-ray luminosity of $\text{L}_{\text{X}}=3 \times 10^{44} \text{erg s}^{-1}$ for energies between 2 and 10 keV \citep{Pounds2001}.
The accretion rate $\dot M$ has a range between 20 and 30\% of the Eddington rate \citep{Petrucci2012} during the XMM-Newton campaign presented in \cite{Kaastra2011}.
The first detection in the 2 to 10 keV band was made with Ariel V \citep{Cooke1978}. The soft-excess was first identified in this object thanks to HEAO1-A2 \citep{Singh1985}; the iron line was for the first time detected in 1987 with EXOSAT \citep{Morini1987} and the reflection component was initially revealed by Ginga \citep{Pounds1994}. EPIC data from XMM-Newton observations (in 2000, 2001, 2005 and 2006) show evidence of a complex Fe K emission line, with a narrow and neutral component possibly produced far from the source, plus a broad and variable component possibly originating in the accretion disk \citep{Pounds2001,Page2003,Ponti2009,Ponti2013}. Using XMM-Newton observations of Mrk 509, \cite{Cappi2009} detected the presence of lines, at $\sim$8-8.5 keV and $\sim$9.7 keV, consistent with being produced by Fe K$\alpha$ and K$\beta$ shell absorptions associated with mildly relativistic and variable outflow. These variable features are not observed during the 2009 campaign \citep{Ponti2013}.

Mrk 509 has been observed during a multi-wavelength campaign presented in \cite{Kaastra2011}, using 5 satellites: XMM-Newton, INTEGRAL, Chandra, HST and Swift, and two ground-based facilities: WHT (optical imaging with Sloan \textit{g}, \textit{r}, \textit{i} and \textit{Z} filters) and PAIRITEL (infrared imaging in \textit{J}, \textit{H} and \textit{K} bands). 
Using the XMM-Newton, Swift, HST and FUSE observations, \cite{Mehdipour2011} showed that the soft X-ray excess varies in association with the thermal optical-UV emission from the accretion disk, suggesting that the soft-excess is due to Comptonization process. 
\cite{Petrucci2012} used the XMM-Newton and INTEGRAL observations of Mrk 509 to model the broad-band emission of this object using physical models, and could explain the entire optical to hard X-ray emission assuming the existence of an accretion disk, a hot, optically-thin corona, responsible of the primary continuum emission, plus a warm ($kT \sim$ 1 keV), optically-thick ($\tau \sim$ 10-20) plasma, creating the optical-UV to soft X-ray emission, plus reflection on distant, neutral material.
As in \cite{Petrucci2012}, we use in this paper the ten simultaneous XMM-Newton (600 ks) and INTEGRAL (1.2 Ms) observations, performed every 4 days in October/November 2009, that enable us to study variability and spectra of Mrk 509 from 0.3 keV up to 200 keV (see Table 1 from \citealt{Kaastra2011}).

%%%%%%%%%%%%%%%%%%%%%%%%%%%%%%%%%%%%%%%%%%%%%%%%%%%%%%%%%%%%%%%%%%%%%%%
\section{Variability spectra}
\label{3}

Taking advantage of the intensive monitoring with XMM-Newton, we first estimate the variability spectra.
We calculate the rms fractional variability amplitude, $\text{F}_{\text{var}}$, following \cite{Vaughan2003}:

\begin{equation}
F_{var}=\sqrt{\frac{S^{2}-\overline{\sigma^{2}_{err}}}{\bar{x}^{2}}}
\end{equation}

with $\bar{x}$ the arithmetic mean of the fluxes,  $\overline{\sigma^{2}_{err}}=\frac{1}{N}\sum_{i=1}^{N}\sigma^{2}_{err,i}$ and $S^{2}=\frac{1}{N-1}\sum_{i=1}^{N}(x_{i}-\bar{x})^{2}$ .

\vspace{2pt}

\begin{figure} [!h]
\resizebox{\hsize}{!}{\includegraphics[trim = 0mm 0mm 0mm 10mm, clip]{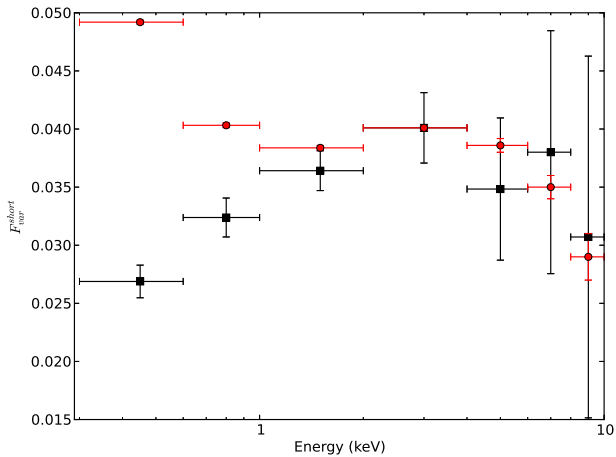}} 
\caption{The black squares represent the variability spectrum on short time scale, $\text{F}_{\text{var}}^{\text{short}}$ ($\text{1 ks} \lesssim \text{T} \lesssim \text{60 ks}$); the red circles are the corresponding variability values predicted from the long time scale variability spectrum, $\text{F}_{\text{var}}^{\text{long}}$ ($\text{60 ks} \lesssim \text{T} \lesssim \text{1 month}$).}
\label{Fvar1}
\end{figure}

We calculate the rms fractional variability on two different time scales, to determine the variability amplitude in Mrk 509 light curves. 
The variability spectrum on time scales 1-60 ks is shown in black squares in Fig. \ref{Fvar1}. We first extract light curves with bin size of 1 ks from XMM-Newton data and calculate rms fractional variability in different energy bands (0.3-0.6 keV, 0.6-1 keV, 1-2 keV, 2-4 keV, 4-6 keV, 6-8 keV and 8-10 keV) for each observation. We then calculate the quadratic mean value over the ten observations to get $\text{F}_{\text{var}}^{\text{short}}$. The error bars represent uncertainties at 1$\sigma$ level.

To obtain the variability on longer time scales, we use the average count rates in the seven energy bins of each observation. We then calculate $\text{F}_{\text{var}}^{\text{long}}$ for all the ten observations. 
As observations are averaged over 60 ks and spread over one month, $\text{F}_{\text{var}}^{\text{long}}$ is sensitive to time scales between $\sim$ 60 ks and $\sim$ 1 month.

If we assume that the observed flux is the sum of a variable component -- the power-law component -- and a less variable component -- the reflection, we can scale the $\text{F}_{\text{var}}^{\text{long}}$ values to the $\text{F}_{\text{var}}^{\text{short}}$ value at 3 keV.
These new values are represented by the red circles in Fig. \ref{Fvar1}.

We can see in Fig. \ref{Fvar1} that, on short time scales (black squares), $\text{F}_{\text{var}}^{\text{short}}$ is maximal between 2 and 4 keV.
The fractional variability decreases in the soft part, and is consistent with a decrease in the hard part. 
Assuming that reflection varies less than the power-law \citep{MiniuttiFabian2004} and that the soft-excess is due to ionized reflection, the variability spectrum is exactly what would be expected, since the variability on short time scales is higher where the power-law dominates.

The shape of the variability spectrum on long time scales (red circles) matches that on short time scales (black squares), but there is a clear excess at low energy.
It shows that the soft-excess is more variable than the power-law on long time scales. 
No variability excess is observed at high energy on long time scales, contrarily to what would be expected if soft-excess and reflection are a single component with no variations in spectral shape (e.g. changes of the ionization parameter should mostly affect the flux in the soft X-ray energy band). The excess of variability in the soft-excess is not compatible with the hypothesis of reflection of hard X-rays, unless the reflection parameters change.

%%%%%%%%%%%%%%%%%%%%%%%%%%%%%%%%%%%%%%%%%%%%%%%%%%%%%%%%%%%%%%%%%%%%%%
\section{Fits with ionized-reflection models}
\label{4}
\subsection{Model}
\label{4.1}

Relativistically blurred ionized reflection has been proposed in several AGN as the origin of the soft-excess. Indeed, iron lines with relativistic profiles have been found in several sources \citep[e.g. in MCG--6--30--15;][]{Tanaka1995}, as well as iron lines with very extended red wings \citep[e.g.][]{Nandra1997}. Disk reflection models use these features to reproduce the emission from a photoionized accretion disk around a supermassive black hole \citep{Ross1993,Ballantyne2001a,RossFabian2005}. These models have been tested successfully in many objects (e.g. on five NLS1 by \citealt{Ballantyne2001b}; on 1H 0707--495 by \citealt{Fabian2002}; on MCG--6--30--15 by \citealt{Ballantyne2003}; and on 34 sources by \citealt{Crummy2006}, using the \textit{reflionx} model from \citealt{RossFabian2005}). Nevertheless, alternative models can also explain the soft-excess in some of these AGN (e.g. warm Comptonization in Ark 564 -- \citealt{Dewangan2007}, and PKS 0558--504 -- \citealt{Papadakis2010}; partial covering in 1H 0707--495 -- \citealt{Tanaka2004}, and MCG-6-30-15 -- \citealt{Miller2008}).

In Mrk 509, the complex iron line at $\sim$6.4 keV is decomposed into a narrow line, constant on a several-year time scale, probably originating from remote neutral material \citep[see][]{DeRosa2004,Ponti2009,Ponti2013} and a resolved component, varying with the continuum on a time scale of a few days \citep{Ponti2013}, that is most probably produced in the broad line region or outer disk. The ionized FeK line (6.7 - 6.97 keV) might be produced in the inner accretion disk.
In order to model the strong soft-excess, we use a \textit{reflionx} component similarly to \cite{Crummy2006}. We convolve the given reflected emission with the Laor model shape \citep[\textit{kdblur};][] {Laor1991}, in order to simulate the blurring from a relativistic accretion disk around a black hole.

In addition to this blurred ionized reflection, we model the power-law continuum produced by Comptonization, including a cut-off at high-energy, using the \textit{cutoffpl} component, as \textit{reflionx} assumes a power-law. To model the narrow component of the iron line, we introduce neutral Compton reflection from distant material (such as the molecular torus postulated in unification models), using the \textit{pexmon} model \citep{Nandra2007}. This model is similar to the \textit{pexrav} model \citep{Magdziarz1995}, but with self-consistent Fe and Ni lines and Compton shoulder \citep{George1991}.
Complex absorption is visible in RGS spectra of Mrk 509 \citep{Detmers2011}.
We therefore add to our model a ``warm absorber" component, because the gas is photoionized in our object.
The final complete model is the following one:\\

\textit{warm absorber (cutoffpl + pexmon + kdblur*reflionx)}\\

The model we apply is a complicated one, with many degrees of freedom, so we put constraints on this model and fix some parameters. 
The warm absorber used for this fitting of Mrk 509 has fixed parameters that have been derived from the 600 ks RGS spectrum \citep{Detmers2011}. It includes all sources of absorption, including the Galaxy. This model of \cite{Detmers2011} has already been used in \cite{Pinto2012} and \cite{Petrucci2012}.
Assuming that neutral reflection originates from distant material, we choose to fix all parameters of the \textit{pexmon} model to the values found in the average spectrum (photon index, reflection factor, abundances and inclination) for the fitting of the ten individual observations. 
The cut-­off energy is fixed to 500 keV, as the \textit{reflionx} model assumes a pure power-­law. This will not affect the modeling using blurred reflection since the cut-­off energy is $>200$ keV \citep{Petrucci2012} during this campaign. In this way, many parameters of \textit{reflionx} are fixed, the free remaining parameters for the individual fits are then the normalization and ionization parameter of the \textit{reflionx} component, the normalization and photon index of the cut-­off power-­law, the index and inner radius of the \textit{kdblur} model.

%%%%%%%%%%%%%%%%%%%%%%%%%%%%%%%%%%%%%%%%%%%%%%%%%%%%%%%%%%%%%%%%%%%%%%%%%%%%%%%%%%%%%%%%
\subsection{Average spectrum}
\label{4.2}

\begin{figure}[!h]
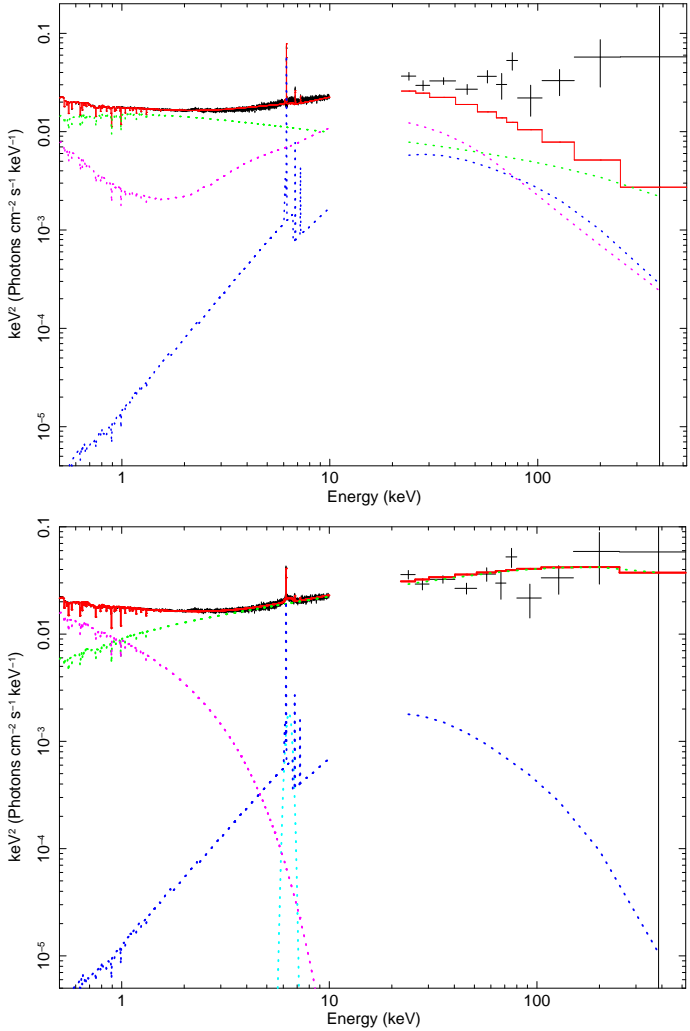

\resizebox{\hsize}{!}{\includegraphics[trim = 0mm 0mm 0mm 0mm, clip, angle=-90]{average.ps}} 
\resizebox{\hsize}{!}{\includegraphics[trim = 0mm 0mm 0mm 0mm, clip, angle=-90]{Comptonization.ps}} 
\caption{Average XMM-Newton and INTEGRAL spectrum of Mrk 509 in XSPEC; Top panel: Data are in black, ionized-blurred reflection fitting result is in red and components of the fitting model are in dash lines (cutoffpl: green, pexmon: blue, kdblur*reflionx: pink); Bottom panel: Data are in black, warm Comptonization fitting result is in red and components of the fitting model are in dash lines (cutoffpl: green, pexmon: dark blue, zgauss: light blue, nthcomp: pink).}
\label{Model}
\end{figure}

In order to test blurred ionized-reflection models to explain the soft-excess in Mrk 509, we fit the average spectrum of the ten XMM-Newton and INTEGRAL observations with the complete model. 
Top panel of Fig. \ref{Model} shows the average spectrum with the resulting fit and the different components of the model, and Table \ref{tableFit} contains the corresponding parameter values. 

\begin{table}
\begin{center}
\caption{Results of the fitting of the average spectrum. Parameters with the symbol * are fixed. Normalizations are given in photons $\text{keV}^{-1} \text{cm}^{-2} \text{s}^{-1}$ at 1keV. Abundances are relative to solar value \protect\citep{Morrison1983}.}
\label{tableFit}
\begin{tabular}{l l l}
 \hline\hline
Model & Parameter & Value\\
  \hline
Cutoffpl -- Reflionx & PhoIndex & 2.24 $\pm$ 0.006 \\
Cutoffpl -- Pexmon & HighECut & 500 keV *\\
Cutoffpl -- Pexmon & Normalization & 1.74 $\pm$ 0.03 $\times 10^{-2}$\\
Pexmon &  PhoIndex & 2.11 $\pm$ 0.01\\
Pexmon &  Rel Refl & 0.79 $\pm$ 0.03\\
Pexmon &  Abund  & 1.00 * \\
Pexmon -- Reflionx &  Fe abund & 1.00 *\\
Pexmon -- Kdblur & Incl & 56.31 $\pm$ 0.45 deg\\
Kdblur &  Index & 10.00 $\pm$ 0.45\\
Kdblur & Rin(G) & 1.235 $\pm$ 0.02 $r_{g}$\\
Kdblur & Rout(G) & 100.00 $r_{g}$ *\\
Reflionx & Xi & 27.27 $\pm$ 1.73 $\text{erg cm s}^{-1}$\\
Reflionx & Normalization & 4.64 $\pm$ 0.75 $\times 10^{-5}$\\
\hline
\end{tabular}
\end{center}
\end{table}

Abundances of iron and other elements are fixed to solar value \citep{Morrison1983}. The fit on the average spectrum gives an inclination angle of the disk of 56 degrees.
The illumination index of the \textit{kdblur} model, characterizing the incident flux distribution, has a very high value of 10. The inner radius also reaches the extreme value of $1.235 ~GM/c^{2}$. These two parameters imply that reflection takes place very close to the central black hole and that this one is rotating with maximal spin. Both these parameters are in fact reaching the model limits of \textit{kdblur} in XSPEC.

\cite{Cerruti2011} tested blurred ionized-reflection models on Mrk 509, using a single epoch Suzaku spectrum. They found that, using \textit{reflext} (an older version of \textit{reflionx}) and \textit{reflionx} models, blurred ionized reflection correctly describes the soft-excess, but underestimates the broad iron emission line, as our model. Resulting parameters of this paper are similar to ours, except the value of the disk emissivity index, equal to 3 in this case. They chose to freeze all \textit{kdblur} parameters to constrain the model and to prevent parameters from reaching the model limits. After freeing the emissivity index parameter, the best-fit value stays consistent with 3. We try to first freeze the emissivity index to 3 and then to let it vary, but even in this case the parameter still reaches the model limit of 10.

We obtain a reduced $\chi^{2}$ around 2.58 for 1340 degrees of freedom, significantly worse than that obtained with a warm Comptonization model. Indeed, by fitting the 0.3-100 keV spectrum with the model used by \cite{Petrucci2012}, but replacing the \textit{compPS} component by a \textit{cutoffpl} model (see bottom panel of Fig. \ref{Model}), we obtain a reduced $\chi^{2}$ around 2.27 for 1338 degrees of freedom (F-test probability equal to $2 \times 10^{-38}$). We note that the INTEGRAL spectrum is not well fitted by ionized-reflection model, contrarily to when a warm Comptonization model is used instead (see bottom panel of Fig. \ref{Model}). Furthermore, the emission is dominated by a very strong reflection, which nevertheless underestimates the power-law-like INTEGRAL spectrum. This bad fit is further discussed in Sect. \ref{6.1}.

\begin{figure}[!h]
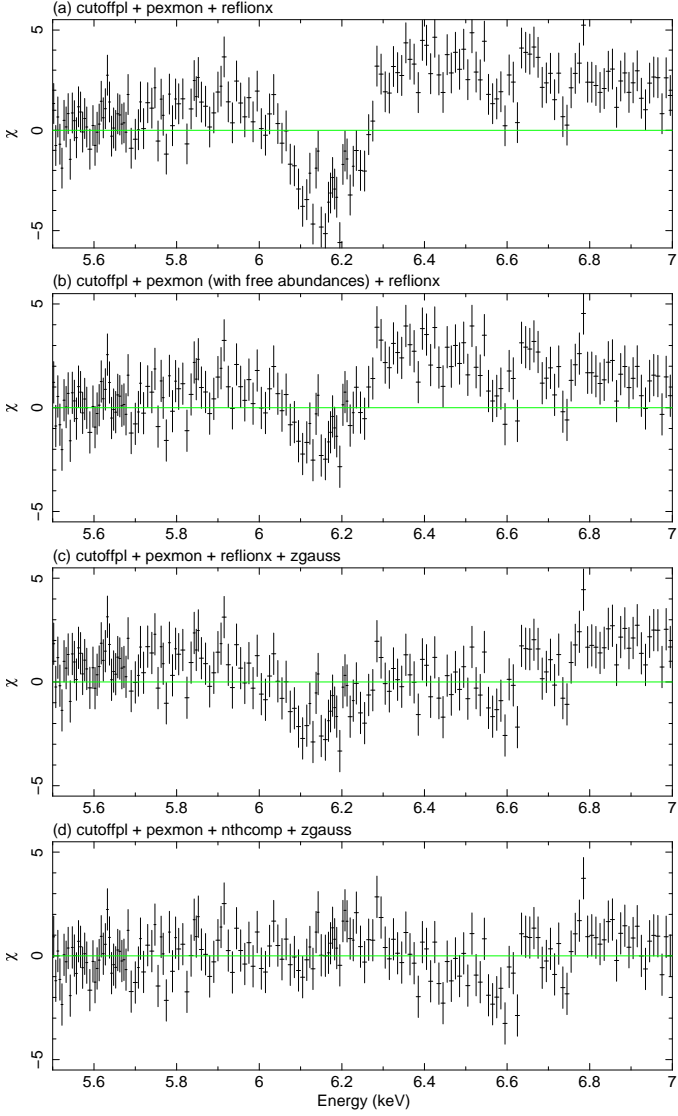

\resizebox{\hsize}{!}{\includegraphics[trim = 0mm 0mm 0mm 0mm, clip, angle=-90]{DelchiIronA.ps}} 
\resizebox{\hsize}{!}{\includegraphics[trim = 0mm 0mm 0mm 0mm, clip, angle=-90]{DelchiIronB.ps}} 
\resizebox{\hsize}{!}{\includegraphics[clip, angle=-90]{DelchiIronC.ps}} 
\resizebox{\hsize}{!}{\includegraphics[trim = 0mm 0mm 0mm 0mm, clip, angle=-90]{DelchiIronD.ps}} 
\caption{Delta $\chi^{2}$ obtained by fitting data with the following models of reflection and Comptonization:
(a) \textit{cutoffpl + pexmon + reflionx}; 
(b) \textit{cutoffpl + pexmon} (with free abundances) \textit{+ reflionx}; 
(c) \textit{cutoffpl + pexmon + reflionx + zgauss}; 
(d) \textit{cutoffpl + pexmon + nthcomp + zgauss}}
\label{ZoomIronDelchi}
\end{figure}

Figure \ref{ZoomIronDelchi} shows fit residuals between 5.5 and 7 keV for different models. Panel (a) shows the ionized-reflection model used in this paper. We note that the iron line and the continuum around the line are not well fitted. 
We obtain a slightly better fit, as shown in panel (b), by letting free the abundances of iron and other elements in the \textit{pexmon} component, but in this case, the iron abundance is very low (28\% of the solar value) and the \textit{pexmon} model dominates the spectrum at high energy. In addition, the fit remains unsatisfactory.
Panel (c) shows the result for the ionized-reflection model to which a Gaussian function has been added in order to represent the broad iron emission line possibly originating from the outer part of the accretion disk, similar to \cite{Petrucci2012}. We fixed the photon index of the \textit{pexmon} to the one of the power-law to prevent the fit from converging on a very hard $\Gamma$=1.4 to better fit the high-energy spectrum. We can see that the fit is better around the iron emission line, but to the detriment of the high energy. However, the fit is still unsatisfactory. A possible explanation is that the \textit{reflionx} model adds a hugely flattened iron line (barely visible in the pink dashed line in Fig. \ref{Model}) that curves the continuum in a way that is incompatible with the data.
Panel (d) shows a similar zoom on the iron line using the model discussed in \cite{Petrucci2012}, which includes a broad iron line to mimic reflection on the accretion disk, but with a single cut-off power-law to represent the continuum, instead of a physical model of Comptonization, for better comparison with the \textit{reflionx} cases. It can be seen that the Comptonization model gives a significantly better fit to the iron line, nearby continuum and also at high energy.

%%%%%%%%%%%%%%%%%%%%%

\subsection{\textit{Reflionx} fits to the individual observations}

As discussed in Sect. \ref{4.1}, we choose to fix all parameters of the \textit{pexmon} model to the values found in the average spectrum and presented in Table \ref{tableFit}. Results of the individual fits are presented in Table \ref{tableFitIndiv}.
By fitting the observations individually, we obtain a mean reduced $\chi^{2}$ around 1.40 for $\sim$510 degrees of freedom. 
The result with a blurred-reflection model is worse than the one obtained with the warm Comptonization model which has a mean reduced $\chi^{2}$ around 1.31 for $\sim$504 degrees of freedom. Note that to be able to compare these two models, we replace the \textit{compPS} component used by \cite{Petrucci2012} to model the continuum by a cut-off power-law \textit{cutoffpl} (parameters of the \textit{Nthcomp} model are similar to those listed in Table 2 from \citealt{Petrucci2012}). 

\begin{table*}
\begin{center}
\caption{Results of the fitting of the average spectrum and each of the ten individual observations. All \textit{pexmon} parameters have been fixed to the values of Table \ref{tableFit}.}
\label{tableFitIndiv}
\begin{tabular}{c c c c c c c c}
 \hline\hline
Spectrum & Photon index & \textit{Cutoffpl} normalization & Index & Rin & $\xi$ & \textit{Reflionx} normalization & $\chi^{2}$/ dof\\
& & ($10^{-2}$ ph. $\text{keV}^{-1} \text{cm}^{-2} \text{s}^{-1}$) & & ($r_{g}$) & $(\text{erg cm s}^{-1})$ & ( $10^{-5}$ ph. $\text{keV}^{-1} \text{cm}^{-2} \text{s}^{-1}$) & \\
  \hline
\hline
Average & 2.24 $\pm$ 0.01 & 1.74 $\pm$ 0.03 & 10.00 $\pm$ 0.45 & 1.235 $\pm$ 0.02 & 27.27 $\pm$ 1.73 & 4.64 $\pm$ 0.75 & 3455/1340 \\
\hline
1 & 2.19 $\pm$ 0.02& 1.51 $\pm$ 0.03 & 10.00 $\pm$ 1.47 & 1.235 $\pm$ 0.09 & 26.24 $\pm$ 2.97 & 3.79 $\pm$ 0.57 & 602/473\\
2 & 2.22 $\pm$ 0.01& 1.64 $\pm$ 0.02 & 10.00 $\pm$ 1.20 & 1.235 $\pm$ 0.08 & 22.12 $\pm$ 1.09 & 5.29 $\pm$ 0.42 & 742/504\\
3 & 2.26 $\pm$ 0.01& 1.88 $\pm$ 0.03 & 10.00 $\pm$ 1.24 & 1.235 $\pm$ 0.08 & 27.09 $\pm$ 3.09 & 5.08 $\pm$ 0.45 & 714/516\\
4 & 2.32 $\pm$ 0.01& 1.46 $\pm$ 0.02 & 10.00 $\pm$ 1.11 & 1.235 $\pm$ 0.07 & 25.17 $\pm$ 1.93 & 5.90 $\pm$ 0.41 & 723/470\\
5 & 2.32 $\pm$ 0.01& 1.78 $\pm$ 0.03 & 10.00 $\pm$ 0.59 & 1.235 $\pm$ 0.07 & 30.48 $\pm$ 3.76 & 5.47 $\pm$ 0.42 & 739/508\\
6 & 2.20 $\pm$ 0.02& 1.84 $\pm$ 0.04 & 10.00 $\pm$ 1.19 & 1.235 $\pm$ 0.08 & 26.87 $\pm$ 3.01 & 5.06 $\pm$ 0.44 & 730/535\\
7 & 2.27 $\pm$ 0.01& 1.89 $\pm$ 0.02 & 10.00 $\pm$ 0.53 & 1.235 $\pm$ 0.08 & 21.75 $\pm$ 0.98 & 6.85 $\pm$ 0.43 & 746/516\\
8 & 2.17 $\pm$ 0.01& 1.91 $\pm$ 0.03 & 10.00 $\pm$ 1.45 & 1.235 $\pm$ 0.09 & 35.78 $\pm$ 4.30 & 3.00 $\pm$ 0.25 & 769/535\\
9 & 2.25 $\pm$ 0.01& 1.69 $\pm$ 0.02 & 10.00 $\pm$ 1.20 & 1.235 $\pm$ 0.08 & 24.53 $\pm$ 1.48 & 5.20 $\pm$ 0.39 & 729/515\\
10 & 2.17 $\pm$ 0.01& 1.77 $\pm$ 0.03 & 10.00 $\pm$ 1.31 & 1.235 $\pm$ 0.08 & 31.05 $\pm$ 4.50 & 3.70 $\pm$ 0.46 & 662/532\\
\hline
\end{tabular}
\end{center}
\end{table*}

Even though the $\chi^{2}$ is significantly better with the Comptonization model, we do not simply discard the blurred-reflection model. Indeed, the model may be over simplified in a number of ways. We will rather focus on the values we find for the fit parameters and on the relationships they have with each other.
We use a simple cut-off power-law model to represent the continuum produced by Comptonization. Unlike \textit{compPS} \citep{Poutanen1996}, used by \cite{Petrucci2012}, which is a physical model of Comptonization, a cut-off power-law is a poor approximation of thermal Comptonization. Nevertheless, we choose \textit{cutoffpl} instead of the \textit{compPS} model because \textit{reflionx} assumes a power-law and because we are not interested in Comptonization details, as the aim of this project is to test ionized-reflection models.

\begin{figure}
		\resizebox{\hsize}{!}{\includegraphics[trim = 0mm 0mm 0mm 10mm, clip]{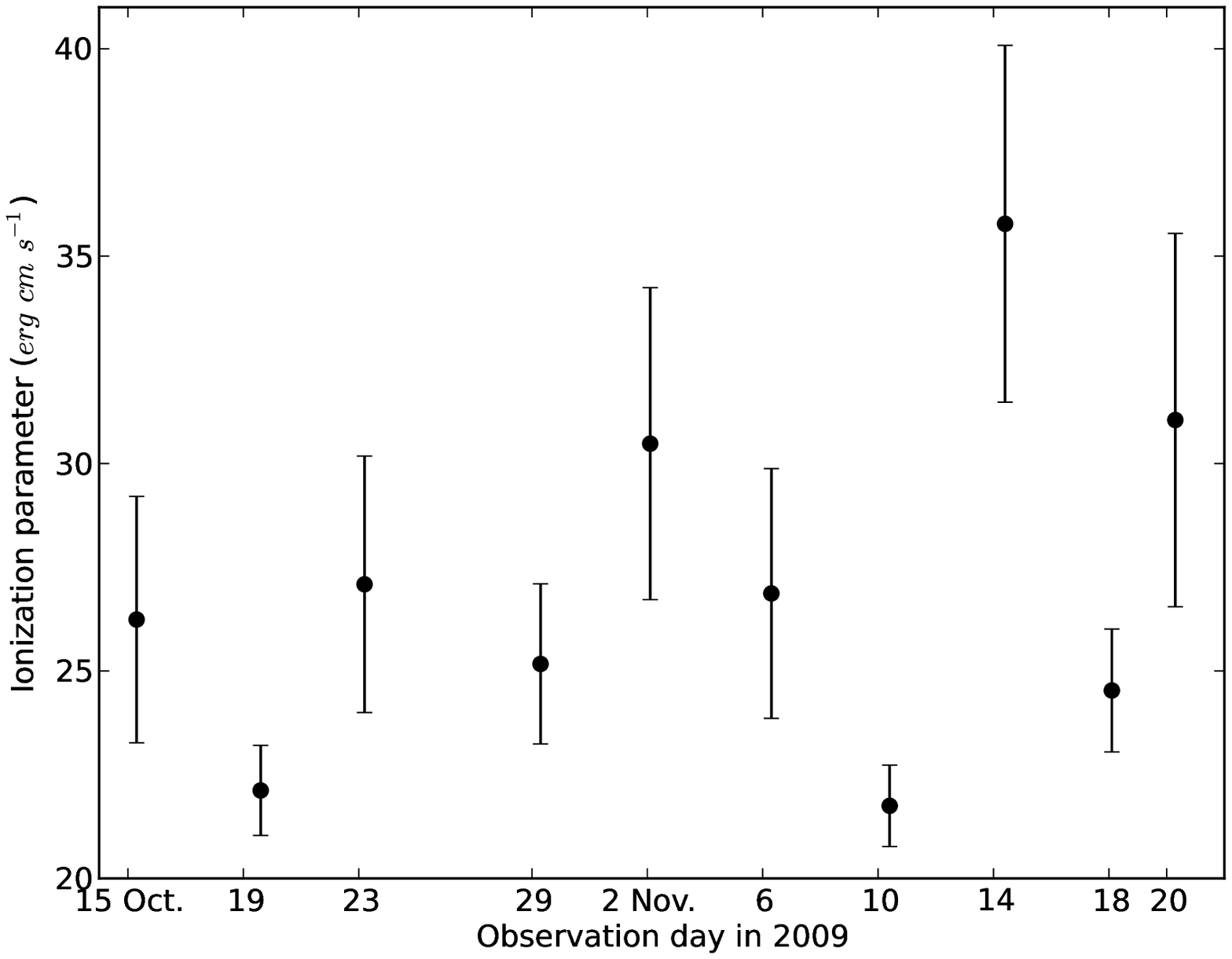}}
		\resizebox{\hsize}{!}{\includegraphics[trim = 0mm 0mm 0mm 10mm, clip]{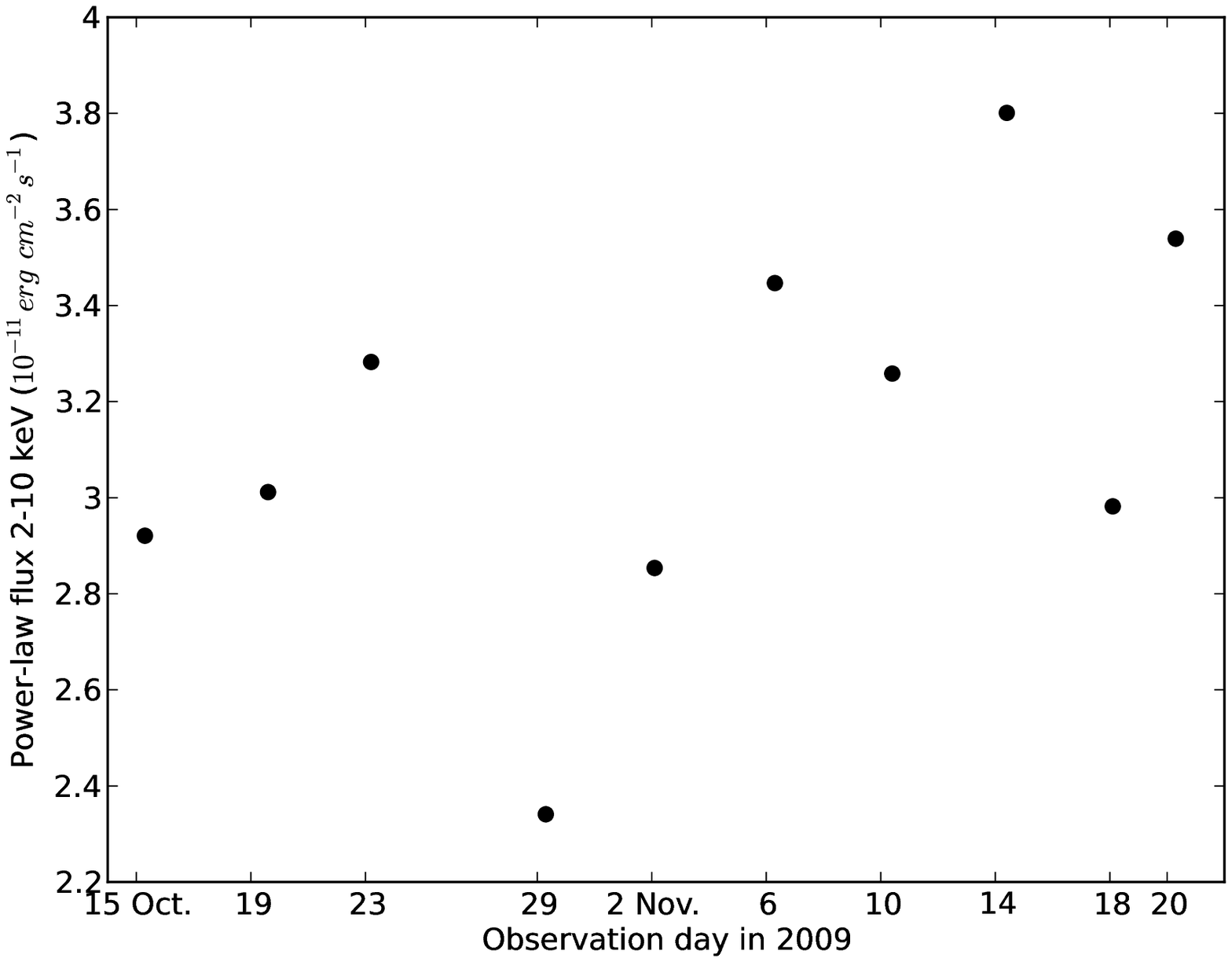}} 
		\resizebox{\hsize}{!}{\includegraphics[trim = 0mm 0mm 0mm 10mm, clip]{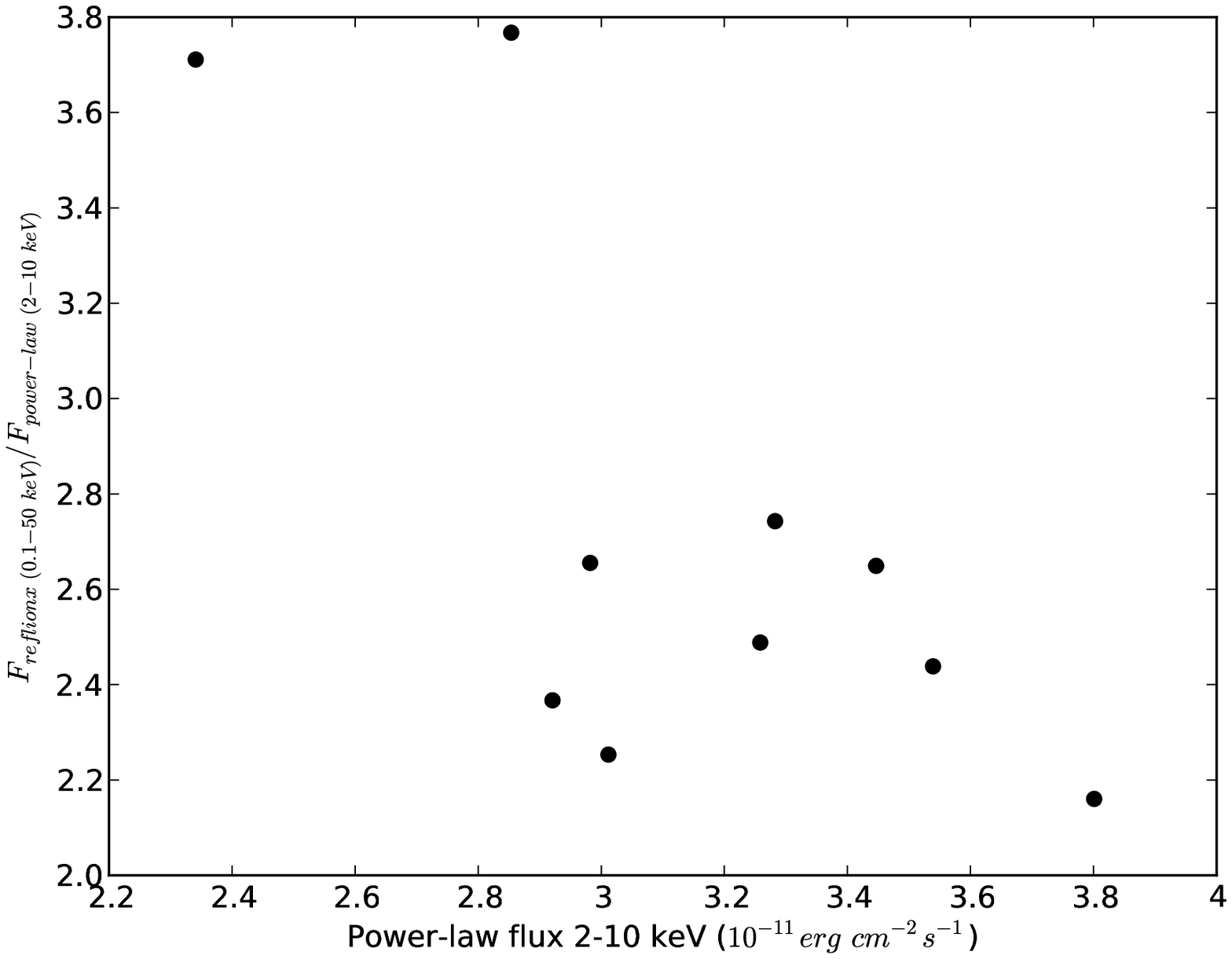}} 
\caption{Top and middle panels: ionization parameter and power-law flux determined for each of the ten observations; Bottom panel: ratio of the reflected flux (0.1-50 keV) over the power-law flux (2-10 keV), as a function of the power-law flux (2-10 keV).}
\label{XiDataObs}
\end{figure}

The emissivity index and the inner radius of the \textit{kdblur} model do not change over the ten observations and keep the extreme values found in the average spectrum.

The ionization parameter $\xi=\frac{L}{nr^{2}} ~\text{erg cm s}^{-1}$, with \textit{L} the luminosity of the ionizing source, \textit{r} the distance of this source to the reflecting medium and \textit{n} the density of the illuminated gas, is in weighted-average equal to 24 $\text{erg cm s}^{-1}$ (see top panel of Fig. \ref{XiDataObs}). 
Such a low value seems at odds with the result that the reflecting medium is located very close to the irradiating source. This is further discussed in Sect. \ref{6.1}.

Middle panel of Fig. \ref{XiDataObs} shows the power-law flux measured between 2 and 10 keV for each observation. The error bars, representing uncertainties at 1$\sigma$ level, are very small. The bottom panel presents the ratio of the reflected flux (0.1-50 keV) over the power-law flux as a function of the continuum power-law flux, showing a significant anti-correlation (Spearman correlation coefficient: r=-0.56; null-hypothesis probability: p=0.09). As for the anti-correlation found, in Fig. 7 from \cite{Fabian2005}, between the reflection factor (i.e. the reflection-dominated component flux relative to the power-law flux) and the continuum flux, our result can be explained by the light-bending model proposed by \cite{MiniuttiFabian2004}. Indeed, in this configuration, the reflection is expected to vary less than the power-law, which is consistent with the behavior of the short time scale variability (black squares in Fig. \ref{Fvar1}), and when the power-law flux drops the spectrum becomes more and more reflection-dominated.

%%%%%%%%%%%%%%%%%%%%%%%%%%%%%%%%%%%%%%%%%%%%%%%%%%%%%%%%%%%%%%%%%%%%%%
\section{Evolution of \textit{reflionx} parameters}
\label{5}

\subsection{Determination of the \textit{R*} parameter for \textit{reflionx}}
\label{chap}

The normalization of the reflected emission component is not related to that of the incident flux, contrarily to most reflection models (\textit{pexrav, pexriv, pexmon}). This makes it not straightforward to determine directly the amplitude of the reflection compared to that of the illuminating flux.
The \textit{pexriv} model describes the reflection of exponentially cut-off power-law spectrum on ionized material \citep{Magdziarz1995}.  
In order to study the evolution of \textit{reflionx} parameters, we first want to determine the equivalent of the reflection factor \textit{R} of \textit{pexriv} for \textit{reflionx}. 
We thus define the pseudo-reflection factor \textit{R*} for \textit{reflionx}. \textit{R*} is a measure of the flux reflected by a reflector with a given ionization, which produces the same observed hard X-ray flux as that predicted with a \textit{pexriv} model with reflection parameter \textit{R}=\textit{R*} ($R=1$ for a reflector with a solid angle of $2\pi$).

In order to determine \textit{R*}, we simulate several \textit{cutoffpl + reflionx} spectra with values of the parameters found in the ten fits with the complete model and fit them with a \textit{pexriv} model. This \textit{pexriv} model contains an explicit reflection factor, an ionization parameter and its normalization is the power-law normalization. 

\textit{Pexriv} does not incorporate the many atomic transitions found at low energy in \textit{reflionx}. However, the high-energy properties (above 20 keV) should be similar to those of \textit{reflionx}. 
The equivalent reflection factor \textit{R*} for \textit{reflionx} is then obtained by fitting \textit{pexriv} to the simulated power-law plus \textit{reflionx} models between 20 and 50 keV, the reflection parameter of \textit{pexriv} giving then the \textit{R*} parameter of \textit{reflionx}.

%%%%%%%%%%%%%%%%%%%%%%%%%%%%%%%%%%%%
\subsection{The case of constant \textit{R*}}
\label{5.2}

We first make the reasonable hypothesis of a stable geometry for Mrk 509, implying a constant reflection factor.
We apply the method described in Sect. \ref{chap} on the average spectrum of the ten observations to determine the reflection factor \textit{R*} and we get $\langle R^{*} \rangle=5.08 \pm 0.82$. 

\begin{figure} [!h]
\resizebox{\hsize}{!}{\includegraphics[trim = 0mm 0mm 0mm 10mm, clip]{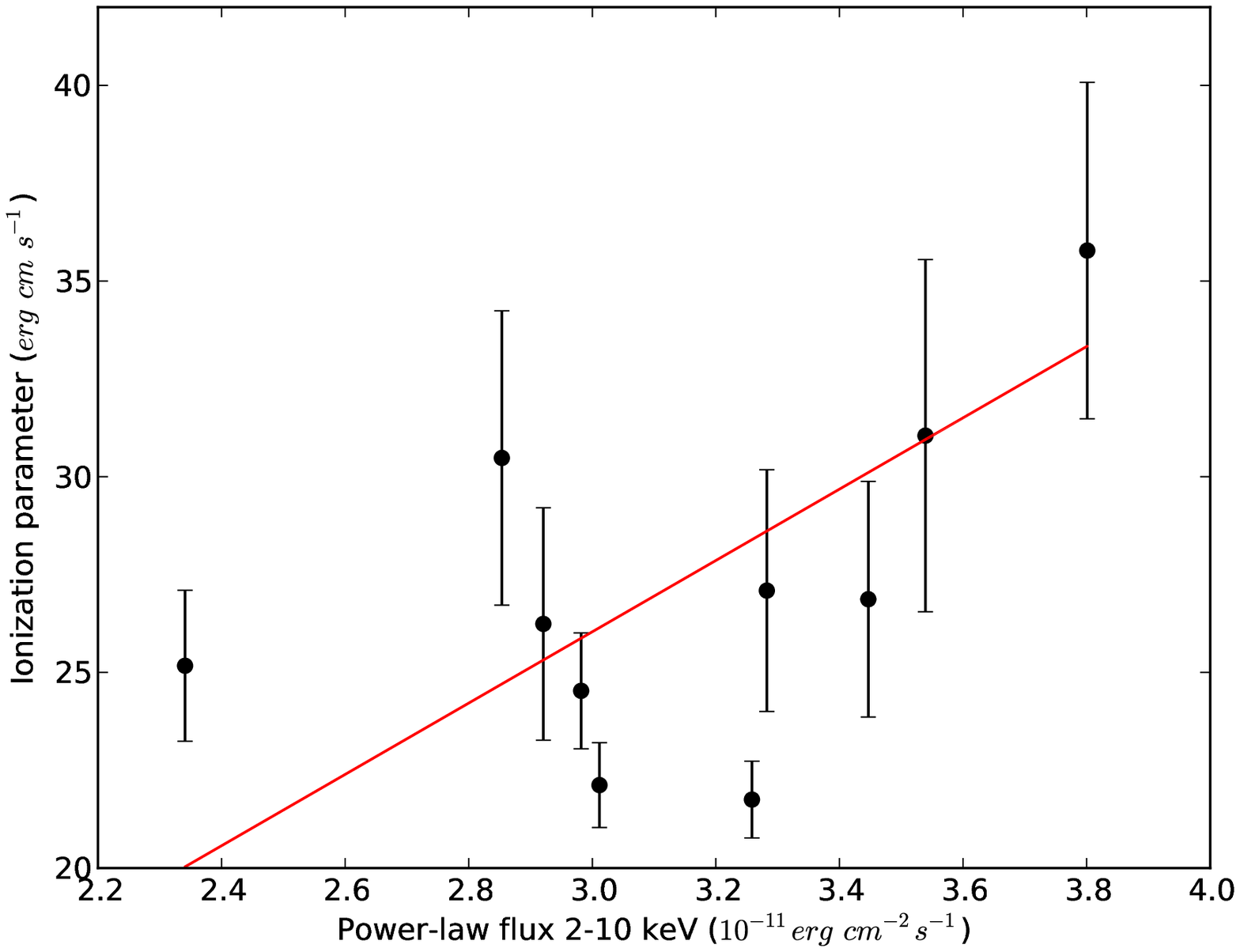}} 
\caption{Ionization parameter as a function of the power-law flux. Red line: evolution expected with fixed \textit{R*}; Black points: parameters obtained in the ten individual fits.}
\label{XiSimuDataFi}
\end{figure}

Figure \ref{XiSimuDataFi} shows the ionization parameter, obtained by fitting each observation with the complete model, as a function of the 2 to 10 keV power-law flux (black points). The red line represents the relation $\xi=\frac{L}{nr^{2}}$ that we expect with our assumption of a constant \textit{R*},
calculated by fixing \textit{R*} to the value of the average spectrum $\langle R^{*} \rangle=5.08$.

The ionization parameter shows a scatter around the expected relation and no significant positive correlation exists (Spearman correlation coefficient: r=0.47; null-hypothesis probability: p=0.17). Taking into account errors in x and y axes, we obtain a chi-squared of $\chi^{2}=73$ for 9 degrees of freedom.

\begin{figure} [!h]
\resizebox{\hsize}{!}{\includegraphics[trim = 0mm 0mm 0mm 10mm, clip]{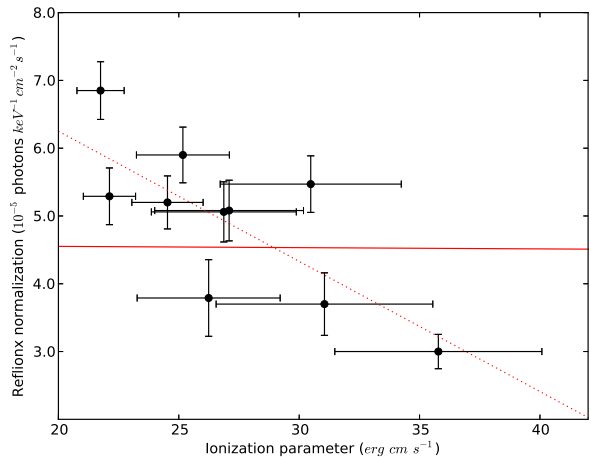}} 
\caption{Reflionx normalization as a function of the ionization parameter. Solid red line: evolution expected with a \textit{reflionx} model; Dot red line: expected evolution of the normalization if the power-law flux is fixed; Black points: parameters obtained by fitting data with the complete model.}
\label{XiSimuDataNormReflionx}
\end{figure}

Figure \ref{XiSimuDataNormReflionx} shows the \textit{reflionx} normalization as a function of the ionization parameter. Black points represent the parameters obtained in the fit of the ten observations. We can see that the \textit{reflionx} normalization decreases significantly with the ionization parameter (Spearman correlation coefficient: r=-0.71; p=0.022). Note that we use the normalization instead of, for instance, the flux, because we are not interested in the physical meaning of parameter, but only in its evolution compared to its expected behavior.

Under the assumption of fixed reflection factor, the ionization parameter and the power-law flux being known, the expected \textit{reflionx} normalization can be deduced. The solid red line in Fig. \ref{XiSimuDataNormReflionx} represents the expected relation between the normalization and $\xi$. The normalization is not expected to vary a lot with the ionization parameter. This is because it should decrease with $\xi$ if the power-law flux is fixed (as shown by the dot red line) and increase with the power-law flux if $\xi$ is fixed. We have seen that $\xi$ increases with the power-law flux, so these two trends are added and compensate each other, inducing a quasi-constant normalization.

Data obviously do not follow the expected behavior ($\chi^{2}=95$ for 10 degrees of freedom, which is significantly worse than the $\chi^{2}$ found in the relation between ionization and power-law flux). This result shows a contradiction with the primary hypothesis of a constant reflection factor. In fact, data seem rather to follow the behavior expected for a fixed power-law flux ($\chi^{2}=12$ for 9 degrees of freedom), even though the power-law flux is strongly variable.

%%%%%%%%%%%%%%%%%%%%%%%%%%%%%%%%%%%
\subsection{Varying \textit{R*}}

\begin{figure} [!h]
		\resizebox{\hsize}{!}{\includegraphics[trim = 0mm 0mm 0mm 10mm, clip]{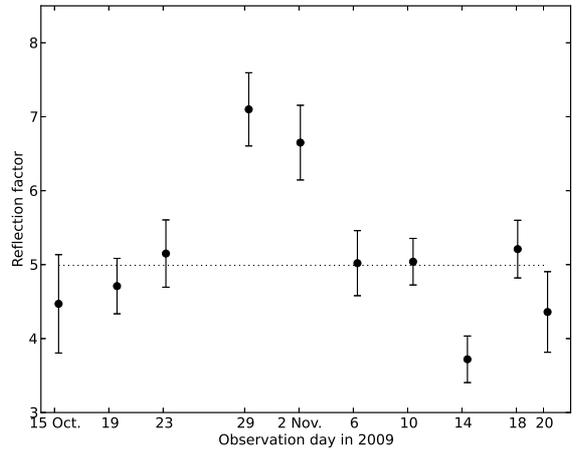}} 
		\resizebox{\hsize}{!}{\includegraphics[trim = 0mm 0mm 0mm 10mm, clip]{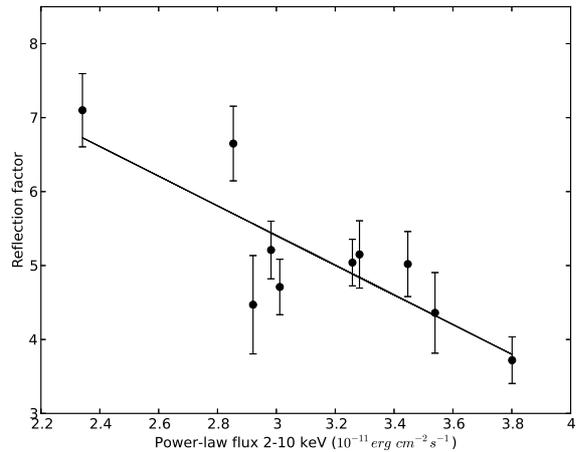}} 
		\resizebox{\hsize}{!}{\includegraphics[trim = 0mm 0mm 0mm 10mm, clip]{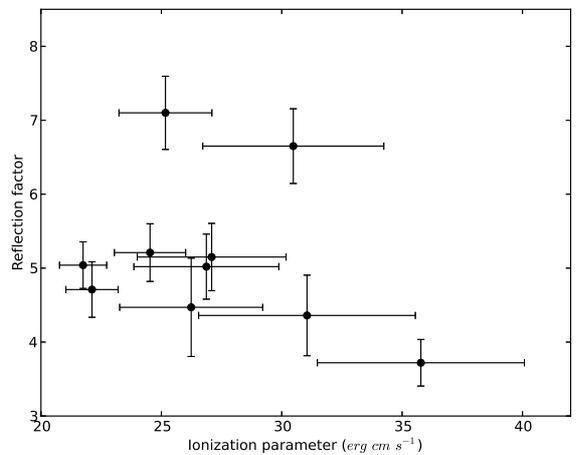}}
\caption{Reflection factor determined for each of the ten observations of Mrk 509 using \textit{reflionx} simulations with parameters values from the complete fitting. Top panel: R* as a function of time; Middle panel: R* as a function of the power-law flux; Bottom panel: R* as a function of the ionization parameter.}
\label{RDataFig}
\end{figure}

We see in Fig. \ref{XiSimuDataNormReflionx} that the reflection factor cannot be constant. We therefore relax the assumption of a constant \textit{R*} and, for each observation, we determine the corresponding \textit{R*}. We use again the method described in Sect. \ref{chap}, using parameters values obtained by fitting data with the complete model. We plot the resulting \textit{R*} as a function of time in the top panel of Fig. \ref{RDataFig}. We can see that the reflection fraction is indeed varying over the observations and that it has a mean value of 4.99. The variability of the reflection factor is characterized by a rms fractional variability amplitude of 0.19 $\pm$ 0.03.

We plot \textit{R*} as a function of the power-law flux in the middle panel of Fig. \ref{RDataFig} and as a function of the ionization parameter in the bottom panel. The reflection factor decreases when the power-law flux increases (Spearman correlation coefficient: r=-0.73; p=0.016). 
This can be explained in the lamp-post model of \cite{MiniuttiFabian2004}, which involves a source located on top of the accretion disk close to the black hole. In this model, changes in source height affect strongly the ratio between the flux emitted towards the observer and the flux emitted towards the accretion disk, because of light-bending effects in the strong gravitational field of the black hole.
However, we do not see any correlation between the reflection factor and the ionization parameter in the bottom panel of Fig. \ref{RDataFig} (r=-0.37; p=0.29). A positive correlation is expected since the reflection factor is a measure of the flux hitting the accretion disk. Thus the pattern of the ionization changes does not match the prediction from the lamp-post configuration.
We can also exclude any change in \textit{R*} originating from changes in the inner radius, since $R_{in}$ is found to remain constant at an extreme value of 1.235 $r_{g}$.

%%%%%%%%%%%%%%%%%%%%%%%%%%%%%%%%%%%%%%%%%%%%%%%%%%%%%%%%%%%%%%%%%%%%%%%
\section{Discussion}
\label{6}

%%%%%%%%%%%%%%%%%%%%%
\subsection{Ionized-reflection model parameters}
\label{6.1}

We tested ionized-reflection models on the ten simultaneous XMM-Newton and INTEGRAL data of Mrk 509 using the \textit{reflionx} XSPEC model. 
While \textit{reflionx} does not provide a better fit to the soft-excess than warm Comptonization \citep{Mehdipour2011,Petrucci2012}, a simple comparison of $\chi^{2}$ is not sufficient to discard this model, because it may be simplified in a number of ways. We discuss here the physical consequences of applying this model to the Mrk 509 data set.

We first notice that the model predicts a very strong reflection (see top panel of Fig. \ref{Model}) and a low ionization, but underestimates the data at high energy.
Such a strong reflection must dominate the hard X-ray spectrum over the power-law, contrarily to what INTEGRAL shows, as it seems power-law dominated. The failure to reproduce the high-energy emission, which is due to the large ionized reflection needed to reproduce the soft-excess is a severe difficulty for blurred ionized reflection models. Two factors may alleviate the difficulty. First, the very low statistical weight of the INTEGRAL data compared to the XMM-Newton ones makes the fit rather insensitive to what happens at high energy, and it is possible that another solution producing a slightly worse overall fit could better reproduce the INTEGRAL data. In addition, we cannot exclude the presence of an additional component, like another, hard, power-law which would start to dominate in the INTEGRAL domain, similar to that observed by \cite{Maiolino2013} in Scorpius X-1, where they found the evidence of a second component of Comptonization up to 200 keV.

Using a warm Comptonization model, it is possible
to have a strong soft-excess without strong reflection, since these two spectral features come from two
different components. Therefore, modeling the soft-excess with thermal Comptonization allows us to obtain a much better representation of the hard X-ray emission (see bottom panel of Fig. \ref{Model}). 

The ionized-reflection model used by \cite{Cerruti2011} gives a good fit for the Suzaku data between 15 and 30 keV. The fit with our ionized-reflection model (see Sect. \ref{4.1}) to the INTEGRAL data does not reproduce the high-energy spectrum as well. We first note that our data extend to much higher energy (20-140 keV). Furthermore, the flux between 17 and 60 keV measured during the INTEGRAL campaign ($6.368 - 6.371 \times 10^{-11} \text{erg} \, \text{cm}^{-2} \text{s}^{-1}$) is significantly higher than the one during the Suzaku observations ($4.81 \times 10^{-11} \text{erg} \, \text{cm}^{-2} \text{s}^{-1}$). \cite{Cerruti2011} also used a two-phase warm absorber, composed by models of partial covering absorption by partially ionized material, instead of the complex three-phase warm absorber derived from RGS data by \cite{Detmers2011}. Doing so, the warm absorber is described with less precision in \cite{Cerruti2011} than in our model, leading to a different soft-excess which results in stronger ionized reflection at high energy.

As the ionization parameter required by the reflection component to fit the 
soft-excess is low, it cannot produce the (possibly relativistic) Fe XXV and 
Fe XXVI lines observed in \cite{Ponti2013}. Instead it will contribute to 
the low ionization emission of the Fe K$\alpha$ line at 6.4 keV. 
The Fe K$\alpha$ line is observed to be composed of a narrow constant component and a resolved and variable component \citep{Ponti2013}, but there is no evidence of relativistically smeared broad iron line in the EPIC spectra. This is also evident in the extreme parameters of \textit{kdblur}, which are effectively reducing the iron line to the absolute minimum, leaving only a very modest bump in the model. Such extreme values of \textit{kdblur} parameters are needed to blur the photoionized emission from the accretion disk in order to reproduce the soft-excess.

We also tried to replace \textit{reflionx} model by a new table model called \textit{xillver} \citep{Garcia2013}, in order to solve the under-prediction of INTEGRAL data. This new version incorporates a richer atomic data base and is expected to provide a more accurate description of the iron Fe K emission line. Unfortunately, using this model did not improve the fit, so we kept the \textit{reflionx} model for our analysis, similarly to \cite{Crummy2006}.

During the fit, abundances of the \textit{pexmon} model have been fixed to solar value, as explained in Sect. \ref{4.2}. We have seen in Fig. \ref{ZoomIronDelchi} that if we let abundances free to vary, we obtain a subsolar iron abundance of 0.28. This value is very problematic as, in the core of a galaxy, we would not expect to have underabundances compared to solar, due to the abundance gradients in galaxies \citep{Diaz1989,Vila1992,Zaritsky1994}. For instance, \cite{Arav2007} found in Mrk 279 absolute abundances of 2-3 times solar.  We also note that \cite{Steenbrugge2011} have found relative abundances consistent with proto-solar abundance ratios in Mrk 509.

The parameters obtained by fitting the data with the ionized-reflection model, such as the illumination index and the inner radius, have extreme values, implying an extreme geometry which maximizes relativistic effects. These parameters are in fact reaching the model limits. 
We also obtain very low ionization values which seem not standard.
Indeed, among the objects analyzed by \cite{Crummy2006}, 12 sources have been fitted with \textit{kdblur} parameters similar to ours, i.e. an illumination index between 8.5 and 10 and an inner radius between 1.2 and 1.4. Only one object has a similar $\xi < \text{50 erg cm s}^{-1}$ ($\xi = \text{30 erg cm s}^{-1}$ in PG 1202+281), while all other objects have $\xi > \text{510 erg cm s}^{-1}$, typical values being around $\text{1600 erg cm s}^{-1}$.

%%%%%%%%%%%%%%%%%%%%%%
\subsection{Hypothesis of a stable geometry}

We first tested the hypothesis of a stable geometry for Mrk 509, implying a constant reflection fraction \textit{R*}, as distances implied are very small and the reflector should react immediately to changes in the continuum. 
With this assumption, some relations are expected between the model parameters. The hypothesis does not hold, as several of the expected relations are not recovered (see Sect. \ref{5.2}). 

The ionization parameter shows a scatter around the expected relation with the power-law flux (see Fig. \ref{XiSimuDataFi}).
The assumption of constant \textit{R*} implies that the \textit{reflionx} normalization is almost constant and independent of the ionization parameter (see Fig. \ref{XiSimuDataNormReflionx}). 
This is due to the fact that, for a fixed $\xi$, the normalization will follow the power-law flux in order to maintain \textit{R*} constant, while, for a fixed power-law flux, the normalization must decrease if the ionization increases.
Both behaviors compensate, which results in a flat relationship. 
Results of fitting of the ten observations with the complete model do not follow this expected trend, showing a clear anti-correlation between the normalization and the ionization, instead of the expected flat relationship. The normalization is behaving as if the power-law flux is fixed, while it is strongly variable (see Fig. \ref{XiDataObs}). 
We conclude that the \textit{reflionx} model, if we assume a stable geometry, adjusts its parameters to compensate changes of one parameter by variations of another one, fine-tuning them in order to fit the data. By doing so, it introduces an anti-correlation between the ionization parameter and the normalization, which is not physically justified.

%%%%%%%%%%%%%%%%%%%%%%
\subsection{Long time scale variability of the soft-excess}

We have seen in Fig. \ref{Fvar1} that, on long time scales (red circles), the variability is maximum where the soft-excess dominates, while on short time scales (black squares) variability is minimum where reflection dominates. 
One possible parameter which could affect differentially soft and hard X-ray variability is the ionization.
Indeed, effects of changes of $\xi$ on the ionized reflection are most important in the energy range where ionized transitions are abundant, i.e. in the soft-excess.
When the ionization state changes, the soft-excess flux varies strongly, which could explain the large variability on long time scale (red circles in Fig. \ref{Fvar1}), assuming that $\xi$ is slowly varying. 

\begin{figure} [!h]
\resizebox{\hsize}{!}{\includegraphics[trim = 0mm 0mm 0mm 10mm, clip]{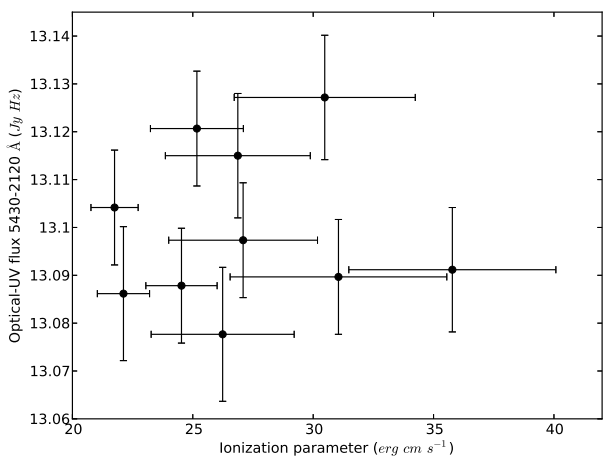}} 
\caption{Optical-UV flux \protect\citep{Mehdipour2011} as function of the ionization parameter estimated in this paper.}
\label{OptiFluxXi}
\end{figure}

The ionization parameter is defined as $\xi=\frac{L}{nr^{2}} ~\text{erg cm s}^{-1}$.
In the case of a stable geometry, \textit{r} is also constant and variations of the ionization parameter can only be due to changes of the density  \textit{n} of the disk, since $\xi$ and $L$ are not well correlated.
We assume that an increase of the UV flux results from an increase of the accretion rate and hence an increase of the density. According to the ionization parameter definition, an increase of the density produces a decrease of the ionization.
Thus the ionization parameter is expected to be anti-correlated with the UV flux. But this behavior is not observed in Fig. \ref{OptiFluxXi}, which shows the optical-UV flux \citep{Mehdipour2011} as a function of the ionization parameter (Spearman correlation coefficient: r=0.18; p=0.63). 
It seems therefore unlikely that the long term variations of the soft-excess are due to changes in ionization state resulting from changes in the density of the reflector.

%%%%%%%%%%%%%%%%%%%%%
\subsection{Lamp-post configuration and light-bending effect}

As the stable geometry with constant reflection factor \textit{R*} is excluded, we drop this hypothesis and allow for a varying geometry.
One such geometry is the lamp-post configuration, with light-bending effects taken into account \citep{MiniuttiFabian2004}. In this configuration, a primary source of X-rays is located close to a central black hole and is illuminating both the observer at infinity and the accretion disk. When the height $h$ of the source above the accretion disk varies, even if the intrinsic luminosity stays constant, the observed flux will vary. The reflection component is shown to vary with much smaller amplitude than the direct continuum \citep{MiniuttiFabian2004}. 
The low variability of the reflection implies that overall flux variability will be maximum where reflection features are weaker, i.e. between 2 and 5 keV.
The variability on short time scale does actually follow the expected behavior from the lamp-post configuration (see Fig. \ref{Fvar1}); however another mechanism is needed to explain the long time scale evolution of the soft-excess.

We calculate  the reflection factor \textit{R*} for each observation (see top panel of Fig. \ref{RDataFig}).
As expected, \textit{R*} is found to be variable, with a mean value of about 5. This high reflection factor again implies a reflection-dominated spectrum in hard X-rays, which is not supported by the INTEGRAL data. Theoretically, the lamp-post configuration allows \textit{R*} values much higher than 1, provided the source is very close to the black hole.
From Fig. 5 in \cite{MiniuttiFabian2004}, \textit{R*} $\sim$ 5 can be obtained if the height of the source above the disk is equal to $h \sim 1 r_{g}$ (when the source is located on the rotation axis) or to $h \sim 2 r_{g}$ (in the case of a corotating source at 2$r_{g}$ from the axis).
These are extreme geometries, implying maximally rotating (or close to maximally rotating) black holes. This conclusion is however in agreement with the extreme parameters of the \textit{kdblur} model.

Middle panel of Fig. \ref{RDataFig} shows that the reflection factor decreases when the power-law flux increases. Bottom panel of Fig. \ref{XiDataObs} also shows an anti-correlation between the ratio of the reflected flux over the power-law flux and the continuum flux, which provides similar information. This behavior is expected in the frame of the lamp-post configuration because, when the source comes closer to the disk, light-bending effect is stronger and induces more photons to be reflected on the accretion disk and less to illuminate the observer at infinity. In this case, we expect to find a correlation between the ionization parameter and the reflection factor because, when the height of the source \textit{h} decreases, more light is reaching the disk, increasing its ionization state. 
However, we find no correlation between \textit{R*} and $\xi$ (see bottom panel of Fig. \ref{RDataFig}). As a consequence, similarly to the case where the reflection factor is assumed to be constant, while the model is able to tune its parameters to fit the individual observations, the expected physical relations between the parameters are not present. This makes us conclude that ionized reflection -- at least in these two configurations -- is unlikely to be the correct explanation for the origin of the soft-excess in Mrk 509.

%%%%%%%%%%%%%%%%%%%%%
\section{Conclusion}
\label{7}

The nature of the soft-excess in AGN is still uncertain. 
While it can be fitted by both ionized reflection \citep{Crummy2006,Zoghbi2008,Fabian2009,Ponti2010} and warm Comptonization \citep{Magdziarz1998,Middleton2009,Jin2009,Mehdipour2011,Petrucci2012} in most, if not all, objects, some detailed features observed in individual objects point either to the former -- lags in the soft-excess \citep{Cackett2013,DeMarco2011,DeMarco2013}; in iron K and L lines \citep{Fabian2009,Zoghbi2013,Kara2013} -- or to the latter -- correlation with UV or variability spectrum \citep{Walter1993,Edelson1996,Marshall1997,Mehdipour2011,Done2012}.
Because blurring can make ionized reflection look very featureless, the distinction between the two models at low energies is difficult and leads to a confusing situation about the origin of the soft-excess. 

In this work we applied the blurred ionized-reflection model in Mrk 509, a bright Seyfert 1 galaxy for which we have a unique data set and which shows the presence of a strong soft-excess. The ionized-reflection model has some difficulty in fitting the broad-band spectrum of Mrk 509, even assuming a very strong reflection and an extreme geometry. We first made the hypothesis of a stable geometry, but this configuration leads to a non-physical anti-correlation between the ionization parameter and the \textit{reflionx} normalization and cannot explain the strong variability of the soft-excess on long time scales. The soft-excess variability cannot be explained by a varying geometry such as a lamp-post configuration either. Furthermore, even if light-bending effects can induce a high reflection factor value, we cannot find the expected correlation between \textit{R*} and $\xi$. The \textit{reflionx} model fine-tunes its parameters in order to fit the data, introducing non-physical relations between parameters and preventing expected relations to appear. Ionized reflection is then unable to explain the origin of the soft-excess in Mrk 509. 

In \cite{Mehdipour2011} and in \cite{Petrucci2012}, the soft-excess of Mrk 509 was attributed to warm Comptonization based on some observed relationships, in particular the correlation between the UV and the soft X-ray fluxes. Under this hypothesis, the excess variability in the soft X-ray flux on long time scales can be explained by changes in the accretion rate (and hence in the seed photon flux). This model of warm Comptonization remains therefore the most probable explanation of the soft-excess in this object.

Ionized reflection is expected as soon as a strong X-ray source irradiates very nearby cold-to-warm matter.
Therefore, it is well possible that both mechanisms are working in all objects, but the dominance of one over the other depends on the physical conditions of the disk and of the corona. 
The advent of sensitive telescopes in the hard X-rays like NuSTAR and in the future ASTRO-H shall provide very useful constraints on the origin of the soft-excess.

One physical parameter that can be expected to determine the existence of blurred ionized reflection is the Eddington ratio. Low Eddington ratios may indeed imply a transition from a standard accretion disk to an advection-dominated accretion flow \citep{Ichimaru1977,Rees1982,Narayan1994,Narayan1995a, Narayan1995b,Abramowicz1995},
effectively truncating the accretion disk and removing cold material from the inner parts where relativistic effects are most important. This is further supported by the fact that the best evidences for blurred ionized reflection, including reverberation lags in the soft-excess \citep{Cackett2013,DeMarco2011} or iron L and K emission \citep{Fabian2009,Zoghbi2013,Kara2013}, are found in Narrow-Line Seyfert 1 objects (for instance 1H 0707--495 and PG 1211+143), which are thought to have very high accretion rates. 
The moderate Eddington ratio of 0.3 in Mrk 509 may therefore be too low for blurred ionized reflection to become the dominant source of soft-excess in this object. However, lags have also been detected in many Seyfert galaxies without high accretion rates \citep{DeMarco2013}.

The ten simultaneous XMM-Newton and INTEGRAL observations of Mrk 509 represent a unique data set which brings unprecedented constraints on the models. 
Many models are able to reproduce the soft-excess, because it is essentially featureless.
Monitoring campaigns allow to study emission models along the time dimension, and can provide very useful constraints to determine the origin of components like the soft-excess. Additional monitoring campaigns on other Seyfert 1 objects, and in particular narrow-line Seyfert 1 objects, where the evidence for ionized reflection is the strongest, would be very useful to understand whether this component can really have different origins in different objects.

%%%%%%%%%%%%%%%%%%%%%%%%%%%%%%%%%%%%%%%%%%%%%%%%%%%%%%%%%%%%%%%%%%%%%%%%%%%%%%%%

\begin{acknowledgements}
This work is based on observations obtained with XMM-Newton, an ESA science mission with instruments and contributions directly funded by ESA Member States and the USA (NASA). It is also based on observations with INTEGRAL, an ESA project with instrument and science data center funded by ESA member states (especially the PI countries: Denmark, France, Germany, Italy, Switzerland, Spain), Czech Republic, and Poland and with the participation of Russia and the USA. RB acknowledges a grant from the Swiss Science National Foundation. GP acknowledges support via an EU Marie Curie Intra-European Fellowship under contract no. FP7-PEOPLE-2012-IEF-331095. P.O. Petrucci acknowledges financial support from the CNES and from the CNRS/PICS-INAF project. We thank the anonymous referee for her/his comments and suggestions.

\end{acknowledgements}

\bibliographystyle{aa}
\bibliography{Biblio.bib}

\end{document}